# Reliability of Ideal Indexes

Gholamreza Hajargasht

The University of Queensland

*Email*: g.hajargasht@uq.edu.au

Abstract

*The Fisher and Gini-Eltetö-Köves-Szulc (GEKS) are celebrated as ideal bilateral and multilateral indexes due to their superior axiomatic and econ-theoretic properties. The Fisher index is the main index used for constructing CPI by statistical agencies and the GEKS is the index used for compiling PPPs in World Bank's International Comparison Program (ICP). Despite such a high status and the importance attached to these indexes, the stochastic approach to these indexes is not well-developed and no measures of reliability exist for these important economic statistics. The main objective of this paper is to fill this gap. We show how appropriate reliability measures for the Fisher and GEKS indexes, and other ideal price index numbers can be derived and how they should be interpreted and used in practice. In an application to 2017 ICP data on a sample of 173 countries, we estimate the Fisher and GEKS indexes along with their reliability measures, make comparisons with several other notable indexes and discuss the implications.*

The author is indebted to D. S. Prasada Rao for helpful suggestions and comments on earlier drafts of this paper. Financial support from the Australian Research Council (ARC) Discovery Project DP170103559 is gratefully acknowledged.

## 1- Introduction

National statistical agencies compile and publish consumer (and producer) price index numbers on a monthly or quarterly basis. Estimates of gross domestic product at constant prices are also published as a part of the national accounts data. International agencies such as the World Bank, OECD and Eurostat produce estimates of purchasing power parities (PPPs) and real expenditures through the International Comparison Program (ICP). A distinguishing feature of the national publications on CPI and international publications on PPPs is the lack of any indication of reliability of the published figures. For example, in the case of international comparisons, one may have a higher level of confidence in price comparisons between countries at a comparable level of development (e.g. between USA and Germany) than a comparison between two "dissimilar" countries (e.g. USA and Kenya). The main objective of this paper is to fill this gap by providing a framework for computing and understanding reliability measures for Fisher, GEKS and other notable index numbers.

Fisher (1922) in his momentous book, *The Making of Index Numbers*, listed 126 index number formulae. Additional formulae have been proposed since then. Among them, the Fisher index has been crowned as the ideal price index for bilateral comparisons due to its superior axiomatic properties and being superlative (see e.g., Consumer Price Index Theory, 2020). It is the index used in constructing consumer price indexes (CPI) by many national and international agencies (also recommended by the Consumer Price Index Manual: Concepts and Methods[1], *2020*) and in constructing purchasing power parities (PPPs) by the World Bank's International Comparison Program (World Bank 2020). The Fisher index is also a building block in constructing multilateral price indexes for temporal and spatial comparisons. Fisher along with Törnqvist are also considered ideal for computing productivity measures both in theory (see e.g., Diewert, 1992) and in practice (e.g., by US Bureau of Labor Statistics). Despite the popularity of the Fisher index, the stochastic approach to it is not well-developed. In particular, no formula for its variance exists.

---
[1] Hereafter we refer to this book as the Manual.

This paper seeks to fill this gap by developing a modern stochastic approach to the Fisher index and provide a simple formula for its variance. We also discuss measures of reliability for other important indexes including Törnqvist, Sato-Vartia and Walsh and compare them to the Fisher index. Empirical results reported here make use of data from the 2017 Round of the International Comparison Program (ICP) covering 173 countries. Other contributions of the paper on bilateral indexes include: (*i*) a stochastic approach and a formula for the variance of the Walsh index; (*ii*) interpreting variance of log-Fisher and other indexes as price dissimilarity measures; and (*iii*) demonstrating that the two main stochastic approaches to index numbers produce the same index and incredibly the same variance.

A bilateral index is suitable for comparing two time periods or two countries. When the comparison is extended to three or more, the index needs to become multilateral. Transitivity is an internal consistency requirement for a multilateral index. None of the widely used index number methods, including Fisher, Tornqvist and Walsh indices, satisfy transitivity. Several approaches exist for obtaining transitive multilateral price indexes (see e.g., Balk 1996; Hill 1997; or Diewert 1999). However, a method due to Gini (1931), Eltetö and Köves (1964) and Szulc (1964), known as the GEKS, is the recommended method for international price comparisons. Since 2005, ICP uses GEKS to compare prices and real incomes across countries (see e.g., Diewert 2013). GEKS is used by the OECD in its triennial comparisons and by Eurostat in its annual comparisons (see Eurostat 2012). GEKS has also been suggested as a good method to construct temporal chained indexes (Diewert and Fox, 2020). Again, despite the importance attached to GEKS, the stochastic approach to GEKS is not well-developed. In this paper we propose an appropriate stochastic approach and use the 2017 ICP data to show that it produces sensible reliability measures and discuss the implications.

The structure of the paper is as follows: Section 2 introduces the new stochastic approach and Section 3 applies it to the Fisher ideal index. Section 4 presents the stochastic approach to several other notable bilateral indexes. In Section 5 we discuss alternative interpretations of the reliability measures. Section 6 develops the stochastic approach to GEKS. In Section 7, we estimate the Fisher and GEKS indexes using 2017 ICP data on a sample of 173 countries and compare them with some of the other notable indexes. Two appendices complement the materials in the text.

### 2- Stochastic Approaches to Price Index Numbers

Suppose there are $N$ commodities indexed with $n=1,....,N$, $p_{nj}$ is the price of $n$-th item in the $j$-th location (or time period)[2] and $e_{nj}$ is expenditure on the $n$-th item in $j$-th location/period. Quantity is defined as $q_{nj}=e_{nj}/p_{nj}$, expenditure shares as $s_{nj} = e_{nj} / \sum_{n=1}^{N} e_{nj}$ and $P_{jk}$ denotes the price index of the *j-th* relative to the *k-th* location/period.

---

[2] - In our notation, $j$ indicates a location (e.g. a country) in the former and a time period in the latter case. We use location/period to include both.

The stochastic approach to index numbers (see e.g. Consumer Price Index Theory, 2020, Chapter 4) can be conducted in two basic ways. One approach relies on modelling *price ratios*, the alternative approach models prices according to the so-called *law of one price*. In the analysis below we use the former approach, but in Appendix-A, we show that the two approaches are equivalent i.e. they produce the same formula for an index and its variance. The stochastic approach to index numbers also requires a way of incorporating weights and an appropriate estimation procedure. There have been two methods to incorporate weights in the stochastic approach. Here we briefly describe these two methods and propose a new method that we use in this paper.

*Weights through heteroskedasticity*: A stochastic approach to index numbers starts with a specification such as

$$\pi_{n,jk} = P_{jk} + v_n \quad n = 1,...,N \tag{1}$$

with $E(v_n) = 0$ and where $\pi_{n,jk}$'s are price relatives (e.g. $\pi_{n,jk} = p_{nj}/p_{nk}$) or log of price relatives and $P_{jk}$ is a bilateral or log of a bilateral price index[3]. Under the early stochastic approach of 1980s and 1990s the weights ($w_n$ s) are incorporated through the variance of the error term by specifying $Var(v_n) = \sigma^2/w_n$ and the model is estimated using generalized least squares (GLS). This heteroskedastic specification has been criticized (see e.g. Diewert 1995) since variability in price ratios may not have a simple relationship with the expenditure share of the commodity. Consequently, this approach has been abandoned.

---

[3] Alternative choices of weights lead to alternative indexes such as Laspeyres, Paasche, Tornqvist, etc.

*Weighted least squares:* In this framework, (1) is still assumed but the weights are incorporated through a weighted least square $Min_{P_{jk}} \sum_{n=1}^{N} w_n (\pi_{n,jk} - P_{jk})^2$ leading to $\hat{P}_{jk} = \sum_{n=1}^{N} w_n \pi_{n,jk} / \sum_{n=1}^{N} w_n$. This approach is often justified by interpreting weights as importance or frequency of price ratios in the population. Note that weights only appear in the estimation stage. The other issue is that sometimes a least square procedure may not be appropriate (see e.g. Appendix-A). It has also been suggested that the weights need to be independent from price ratios for this method to be valid (e.g. by Gorajek 2018) but we are not sure if this is the case.

*Weighted method of moments*: Since index numbers are basically averages of price relatives, a third estimation procedure can be developed entirely based on moments. This approach starts by defining an index for the population as a weighted mean. Let $w_n$ and $\pi_{n,jk}$ have a joint density $f_n(w_n, \pi_{n,jk})$, then the index for population is defined as[4]

$$P_{jk} = E(w_n \pi_{n,jk}) / E(w_n) \qquad (2)$$

with $E(w_n \pi_{n,jk}) = \iint w_n \pi_{n,jk} f_n(w_n, \pi_{n,jk}) dw_n d\pi_{n,jk}$, $E(w_n) = \int w_n g_n(w_n) dw_n$ and $g_n(w)$ is the density function of $w_n$. (2) can also be written as

$$E\{w_n (\pi_{jk} - P_{jk})\} = 0 \qquad (3)$$

then $P_{jk}$ can be estimated by solving the sample moment analogue of (3)

---

[4] One can start with a less restrictive definition $P_{jk}^L = E\left(\sum_{n=1}^{N} w_n \pi_{n,jk}\right) / E\left(\sum_{n=1}^{N} w_n\right)$ which gives the same index but not necessarily the same variance. We leave this framework for future studies.

$$\frac{1}{N}\sum_{n=1}^{N} w_n (\pi_{n,jk} - P_{jk}) = 0 \Rightarrow P_{jk} = \sum_{n=1}^{N} w_n \pi_{n,jk} \Big/ \sum_{n=1}^{N} w_n \qquad (4)$$

Note that this estimation method relies only on the definition of the index i.e. (2), it does not require the specific form (1). Also, the weights are not required to be uncorrelated with price ratios.

The method of moments can also be used to obtain variances. Suppose there are $N$ observations, $\mathbf{r}(\boldsymbol{\theta})$ is a $K \times 1$ vector of moment conditions, $\boldsymbol{\theta}$ is a $K \times 1$ vector of parameters[5] to be estimated and we have the moment conditions

$$E[\mathbf{r}_n(\boldsymbol{\theta})] = 0 \qquad (5)$$

Then $\boldsymbol{\theta}$ can be estimated by solving $\bar{\mathbf{r}}(\hat{\boldsymbol{\theta}}) = \frac{1}{N}\sum_{n=1}^{N} \mathbf{r}_n(\hat{\boldsymbol{\theta}}) = 0$ and under certain regularity conditions, the resulting $\hat{\boldsymbol{\theta}}$ is consistent and asymptotically normal with variance

$$Var(\hat{\boldsymbol{\theta}}) = \left(\frac{\partial \bar{\mathbf{r}}'}{\partial \hat{\boldsymbol{\theta}}}\right)^{-1} Var(\bar{\mathbf{r}}) \left(\frac{\partial \bar{\mathbf{r}}'}{\partial \hat{\boldsymbol{\theta}}}\right)^{-1} \qquad (6)$$

It is often assumed that the observations are independent across $n$ under which $Var(\bar{r}_j) = \frac{1}{N^2}\sum_{n=1}^{N} r_j(\hat{\boldsymbol{\theta}})^2$. We can however allow correlations across moment conditions for each $n$ leading to $Cov(\bar{r}_j, \bar{r}_k) = \frac{1}{N^2}\sum_{n=1}^{N} r_{j,n}(\hat{\boldsymbol{\theta}}) r_{k,n}(\hat{\boldsymbol{\theta}})$.

Note that the weighted least square and our method of moments often produce the same index and variance. So, in most cases it does not matter which approach to use. There are however situations under which a least square procedure does not provide a consistent estimator.

---

[5] $\mathbf{r}$ and $\boldsymbol{\theta}$ must have the same dimensions.

### 3- Stochastic Approach to the Fisher index

The *Fisher Index* is defined as the geometric mean of the Laspeyres and Paasche indexes. *Laspeyres index* of *j* with *k* as the base is defined by

$$P_{jk}^L = \frac{\sum_{n=1}^{N} p_{nj} q_{nk}}{\sum_{n=1}^{N} p_{nk} q_{nk}} = \sum_{n=1}^{N} s_{nk} \frac{p_{nj}}{p_{nk}} \qquad (7)$$

The *Paasche Index* is given by

$$P_{jk}^P = \frac{\sum_{n=1}^{N} p_{nj} q_{nj}}{\sum_{n=1}^{N} p_{nk} q_{nj}} = \left( \sum_{n=1}^{N} s_{nj} \frac{p_{nk}}{p_{nj}} \right)^{-1} \qquad (8)$$

Defining $r_{L,n} = s_{nk} \left( p_{nj}/p_{nk} - P_{jk}^L \right)$ and $r_{P,n} = s_{nj} \left( p_{nk}/p_{nj} - 1/P_{jk}^P \right)$, a stochastic approach to the Fisher index can be initialized by writing moment conditions

$$E(r) = \begin{bmatrix} E(r_{L,n}) \\ E(r_{P,n}) \end{bmatrix} = \begin{bmatrix} E\{s_{nk}(p_{nj}/p_{nk} - P_{jk}^L)\} \\ E\{s_{nj}(p_{nk}/p_{nj} - 1/P_{jk}^P)\} \end{bmatrix} = \begin{bmatrix} 0 \\ 0 \end{bmatrix} \qquad (9)$$

Solving the sample analogue

$$\bar{r} = \begin{bmatrix} \bar{r}_L \\ \bar{r}_P \end{bmatrix} = \begin{bmatrix} \frac{1}{N} \sum_{n=1}^{N} s_{nk} \left( p_{nj}/p_{nk} - P_{jk}^L \right) \\ \frac{1}{N} \sum_{n=1}^{N} s_{nj} \left( p_{nk}/p_{nj} - 1/P_{jk}^P \right) \end{bmatrix} = \begin{bmatrix} 0 \\ 0 \end{bmatrix} \qquad (10)$$

gives the Laspeyres and Paasche indexes (7) and (8). To obtain variances, note that the two equations in (9) are highly correlated for each *n* and as a result we need to allow for correlation. A reasonable correlation structure can be specified as

$$\Omega = \begin{bmatrix} \sigma_1^2 & 0 & & \varsigma_1 & 0 & \\ 0 & \ddots & & & 0 & \ddots \\ & & \sigma_N^2 & & & \varsigma_N \\ \varsigma_1 & 0 & & \tau_1^2 & 0 & \\ 0 & \ddots & & & 0 & \ddots \\ & & \varsigma_N & & & \tau_N^2 \end{bmatrix} \quad (11)$$

where correlation is non-zero across identical items but zero across non-identical items, $\sigma_n^2 = Var(r_{L,n})$, $\tau_n^2 = Var(r_{P,n})$ and $\varsigma_n = cov(r_{L,n}, r_{P,n})$. Then using (6)

$$Var\begin{pmatrix} P_{jk}^L \\ 1/P_{jk}^P \end{pmatrix} = \left(\frac{\partial \bar{r}}{\partial [P_{jk}^L \ P_{jk}^L]}\right)^{-1} Var(\bar{r}) \left(\frac{\partial \bar{r}}{\partial [P_{jk}^L \ P_{jk}^L]}\right)^{-1}$$

with $\left(\frac{\partial \bar{r}}{\partial [P_{jk}^L \ P_{jk}^L]}\right)^{-1} = \begin{bmatrix} \frac{\partial \bar{r}_L}{\partial P_{jk}^L} & \frac{\partial \bar{r}_L}{\partial (1/P_{jk}^P)} \\ \frac{\partial \bar{r}_P}{\partial (\partial P_{jk}^L)} & \frac{\partial \bar{r}_P}{\partial (1/P_{jk}^P)} \end{bmatrix}^{-1} = \begin{bmatrix} N & 0 \\ 0 & N \end{bmatrix}$ and $Var(\bar{r}) = \frac{1}{N^2} \begin{bmatrix} \sum_{n=1}^{N} \hat{\sigma}_n^2 & \sum_{n=1}^{N} \hat{\sigma}_n \hat{\tau}_n \\ \sum_{n=1}^{N} \hat{\sigma}_n \hat{\tau}_n & \sum_{n=1}^{N} \hat{\tau}_n^2 \end{bmatrix}$

which leads to

$$Var\begin{pmatrix} P_{jk}^L \\ 1/P_{jk}^P \end{pmatrix} = \begin{bmatrix} \sum_{n=1}^{N} \hat{\sigma}_n^2 & \sum_{n=1}^{N} \hat{\sigma}_n \hat{\tau}_n \\ \sum_{n=1}^{N} \hat{\sigma}_n \hat{\tau}_n & \sum_{n=1}^{N} \hat{\tau}_n^2 \end{bmatrix} \quad (12)$$

where $\hat{\sigma}_n = s_{nk}\left(\frac{p_{nj}}{p_{nk}} - P_{jk}^L\right)$ and $\hat{\tau}_n = s_{nj}\left(\frac{p_{nk}}{p_{nk}} - \frac{1}{P_{jk}^P}\right)$.

It is often more useful to derive the formula for variance of the logarithm of an index. A simple application of the *delta method*[6] gives:

$$Var(\ln P_{jk}^L) = \sum_{n=1}^{N} s_{nk}^2 \left\{\frac{p_{nj}}{p_{nk} P_{jk}^L} - 1\right\}^2$$

---

[6] According to the delta method if $\hat{\theta}$ is asymptotically distributed as $N(\mathbf{0}, \hat{\mathbf{V}})$ then $Var[f(\hat{\theta})] = \frac{\partial f(\hat{\theta})}{\partial \hat{\theta}}' \hat{\mathbf{V}} \frac{\partial f(\hat{\theta})}{\partial \hat{\theta}}$.

$$Var\left(\ln\left[1/P_{jk}^{P}\right]\right) = \sum_{n=1}^{N} s_{nj}^{2} \left\{ \frac{p_{nk} P_{jk}^{P}}{p_{nj}} - 1 \right\}^{2} \tag{13}$$

$$Cov\left(\ln P_{jk}^{L}, \ln\left[1/P_{jk}^{P}\right]\right) = \sum_{n=1}^{N} s_{nk} s_{nj} \left\{ \frac{p_{nj}}{p_{nk} P_{jk}^{L}} - 1 \right\} \left\{ \frac{p_{nk} P_{jk}^{P}}{p_{nj}} - 1 \right\}$$

To obtain the Fisher index, we first note that $P_{jk}^{F} = \sqrt{P_{jk}^{L} P_{jk}^{P}}$ or in logarithmic form

$$\ln(P_{jk}^{F}) = \frac{1}{2}\left(\ln P_{jk}^{L} - \ln(1/P_{jk}^{P})\right) \tag{14}$$

Therefore $Var\left(\ln P_{jk}^{F}\right) = \frac{1}{4}\left\{Var\left(\ln P_{jk}^{L}\right) + Var\left(\ln\left[1/P_{jk}^{P}\right]\right) - 2Cov\left(\ln P_{jk}^{L}, \ln\left[1/P_{jk}^{P}\right]\right)\right\}$ (15)

which can be simplified to

$$Var\left(\ln P_{jk}^{F}\right) = \frac{1}{4}\sum_{n=1}^{N}\left\{s_{nk}\left(\frac{p_{nj}}{p_{nk} P_{jk}^{L}} - 1\right) - s_{nj}\left(\frac{p_{nk} P_{jk}^{P}}{p_{nj}} - 1\right)\right\}^{2} \tag{16}$$

This is our simple formula for variance of the logarithm of the Fisher index. If the variance of the Fisher Index itself is desired, once again the *delta method* can be applied to obtain[7]:

$$Var\left(P_{jk}^{F}\right) = \frac{1}{4}\sum_{n=1}^{N}\left\{s_{nk}\left(\frac{p_{nj}}{p_{nk} P_{jk}^{L}} - 1\right) - s_{nj}\left(\frac{p_{nk} P_{jk}^{P}}{p_{nj}} - 1\right)\right\}^{2} P_{jk}^{F^{2}} \tag{17}$$

The only other study that has tried to derive a formula for variance of the logarithm of the Fisher index that we are aware of is Cuthbert (2003). He makes different assumptions about the underlying stochastic process and arrives at a much more complicated formula. He assumes that measurement of prices and expenditures are prone to errors through equations $p_{nk} = p_{nk}^{0}(1+v_{it})$ and $e_{nk} = e_{nk}^{0}(1+u_{it})$ respectively

---

[7] According to the method of moments and the delta method, the Fisher index and its log are asymptotically normal with variances given by (16) and (17) respectively.

where $v_{it}$ and $u_{it}$ are random errors with unknown homoscedastic variances $\sigma_p^2$ and $\sigma_e^2$ and where $p_{nk}^0$ and $e_{nk}^0$ are unknown "true" measures of prices and expenditures. The main issue with Cultbert's formula is that it depends on unknown parameters $\sigma_p^2$ and $\sigma_e^2$ where assumptions need to be made about their values. The estimated standard errors based on Cuthbert's formula are often too small (see e.g., Table 1A in Cuthbert 2003). We think our assumptions are more natural leading to the simple and intuitive formula (16). As we will see later in Section 7, standard errors based on our formula are almost identical to those from the nonparametric bootstrap which is not based on any particular stochastic specification. These results confirm that our formula is robust and indeed measures the uncertainty around the index.

### 4- Stochastic Approach to some other Notable Indexes

There are several other notable indexes, among them are the Törnqvist, Sato-Vartia and Walsh. Both the economic and the test approaches to index numbers indicate excellent properties for these indexes (see e.g. Balk 2008 or the Manual 2020).

*4-1 Description of the Indexes*

*Törnqvist:* The Törnqvist (Törnqvist, 1936) index is widely used and one of the recommended indexes in the Manual (2020) due to its excellent axiomatic and economic properties. The US Bureau of Labor Statistics uses the Törnqvist price index as its target index for its chained Consumer Price Index (CPI). The stochastic approach to the Törnqvist index is well-developed going back to Theil (1967; 136-137). It is defined as a weighted geometric mean of price ratios:

$$P_{jk}^T = \prod_{n=1}^{N}\left(\frac{p_{nj}}{p_{nk}}\right)^{\omega_n^A} \tag{18}$$

where $\omega_n^A = \dfrac{s_{nj} + s_{nk}}{2}$ is the average of the expenditure shares in periods *j* and *k*.

*Sato-Vartia*: The Sato (1976)–Vartia (1976) (SV) index is another logarithmic bilateral index which satisfies factor reversal test[8]. It is the first log-change index found to satisfy all the test properties of the Fisher index. The SV index is an exact index for constant elasticity of substitution (CES) utility function (Sato, 1976; Feenstra, 1994). It features prominently in trade and macroeconomic analyses (Feenstra, 1994; Redding and Weinstein, 2020; Balk et al., 2020). It is defined as

$$P_{jk}^{SV} = \prod_{n=1}^{N}\left(\frac{p_{nj}}{p_{nk}}\right)^{\omega_n^L} \text{ with } \omega_n^L = \frac{L(s_{nj}, s_{nk})}{\sum_{n=1}^{N} L(s_{nj}, s_{nk})} \tag{19}$$

where $L(s_{nj}, s_{nk})$ is logarithmic mean of $s_{nj}$ and $s_{nk}$ i.e. $L(s_{nj}, s_{nk}) = \dfrac{s_{nj} - s_{nk}}{\ln s_{nj} - \ln s_{nk}}$.

*Bilateral Product Dummy Index:* Another related logarithmic bilateral index is the so-called product dummy (PD) index[9]. The product dummy method is often used for constructing multilateral indexes (see e.g., Rao 2005) but it can also be used to derive a binary index. The binary index based on the product dummy method turns into a logarithmic index similar to the Törnqvist with the difference that the weights are the harmonic means of the expenditure shares (see e.g., Rao 2005 and Diewert 2005):

$$P_{jk}^{PD} = \prod_{n=1}^{N}\left(\frac{p_{nj}}{p_{nk}}\right)^{\omega_n^H} \text{ with } \omega_n^H = \frac{H(s_{nj}, s_{nk})}{\sum_{n=1}^{N} H(s_{nj}, s_{nk})} \tag{20}$$

where $H(s_{nj}, s_{nk})$ denotes the harmonic mean of $s_{nj}$ and $s_{nk}$.

---

[8] The Törnqvist–Theil index fails the factor reversal test.
[9] The temporal version of this index is known as TPD (Time Product Dummy) and its spatial version as CPD (Country Product Dummy).

*Walsh Index:* Another index which is an ideal index under an alternative axiomatic and economic framework is the Walsh index. It is a superlative index and one of the recommended indexes by the Manual (2020). The Walsh index is defined by

$$P_{jk}^W = \frac{\sum_{n=1}^{N}\sqrt{q_{nj}q_{nk}}\,p_{nj}}{\sum_{n=1}^{N}\sqrt{q_{nj}q_{nk}}\,p_{nk}} = \frac{\sum_{n=1}^{N}\sqrt{s_{nj}s_{nk}}\left(p_{nj}/p_{nk}\right)^{1/2}}{\sum_{n=1}^{N}\sqrt{s_{nj}s_{nk}}\left(p_{nk}/p_{nj}\right)^{1/2}} \qquad (21)$$

## 4-2 Variance of the Indexes

*Törnqvist, Bilateral PD and Sato-Vartia:* Consider the following general moment condition where $\omega_n$ represents a set of weights.

$$E(r_n) = E\left\{\omega_n\left(\ln\left(p_{nj}/p_{nk}\right) - \ln P_{jk}\right)\right\} = 0 \qquad (22)$$

Solving the sample moment conditions leads to

$$\ln P_{jk} = \sum_{n=1}^{N}\omega_n \ln \frac{p_{nj}}{p_{nk}} \qquad (23)$$

If $\omega_n = \omega_n^A$ we obtain Törnqvist, if $\omega_n = \omega_n^H$ we obtain bilateral PD and if $\omega_n = \omega_n^{SV}$ we obtain the Sato-Vartia index. The estimated variance of $\ln P_{jk}$ is given by

$$Var(\ln P_{jk}) = \left(\frac{\partial \bar{r}}{\partial \ln P_j^k}\right)^{-1} Var(\bar{r}) \left(\frac{\partial \bar{r}}{\partial \ln P_j^k}\right)^{-1} \qquad (24)$$

with $\dfrac{\partial \bar{r}}{\partial \ln P_j^k} = \dfrac{1}{N}\sum_{n=1}^{N}\omega_n = \dfrac{1}{N}$ and $Var(\bar{r}_L) = \dfrac{1}{N^2}\sum_{n=1}^{N}s_{nk}^2\left\{\dfrac{p_{nj}}{p_{nk}} - P_{jk}^L\right\}^2$ leading to

$$Var(\ln P_{jk}) = \sum_{n=1}^{N}\omega_n^2\left\{\ln\left(\frac{p_{nj}}{p_{nk}}\right) - \ln P_{jk}\right\}^2 \qquad (25)$$

*Walsh Index:* We follow a similar strategy that we used for the Fisher index to obtain the variance formula for the Walsh index. First, let's define $P_{jk}^W = P_{jk}^a / P_{jk}^b$ where

$$P_{jk}^a = \sum_{n=1}^{N} \varpi_n \left( p_{nj} / p_{nk} \right)^{1/2}, \quad P_{jk}^b = \sum_{n=1}^{N} \varpi_n \left( p_{nk} / p_{nj} \right)^{1/2} \text{ and write}$$

$$\begin{bmatrix} E(r_{1,n}) \\ E(r_{2,n}) \end{bmatrix} = \begin{bmatrix} E\left\{ \varpi_n \left[ \left( p_{nj} / p_{nk} \right)^{1/2} - P_{jk}^a \right] \right\} \\ E\left\{ \varpi_n \left[ \left( p_{nk} / p_{nj} \right)^{1/2} - P_{jk}^b \right] \right\} \end{bmatrix} = \begin{bmatrix} 0 \\ 0 \end{bmatrix} \quad (26)$$

where $\varpi_n = \sqrt{s_{nj} s_{nk}} \Big/ \sum_{n=1}^{N} \sqrt{s_{nj} s_{nk}}$. Solving the sample moment analogue

$$\bar{r} = \begin{bmatrix} \bar{r}_1 \\ \bar{r}_2 \end{bmatrix} = \begin{bmatrix} \dfrac{1}{N} \sum_{n=1}^{N} \left\{ \varpi_n \left( \left( p_{nj} / p_{nk} \right)^{1/2} - P_{jk}^a \right) \right\} \\ \dfrac{1}{N} \sum_{n=1}^{N} \left\{ \varpi_n \left( \left( p_{nk} / p_{nj} \right)^{1/2} - P_{jk}^b \right) \right\} \end{bmatrix} = \begin{bmatrix} 0 \\ 0 \end{bmatrix} \quad (27)$$

gives $P_{jk}^a$ and $P_{jk}^b$ defined earlier. Again, the two equations in (26) are highly correlated for each $n$ and as a result we need to specify a correlation structure such as (11) where $\sigma_n^2 = Var(r_{1,n})$, $\tau_n^2 = Var(r_{2,n})$ and $\varsigma_n = \text{cov}(r_{1,n}, r_{2,n})$. To derive the variance, we again use

$$Var\begin{pmatrix} P_{jk}^a \\ P_{jk}^b \end{pmatrix} = \left( \frac{\partial \bar{r}}{[\partial P_{jk}^a \ \partial P_{jk}^b]} \right)^{-1} Var(\bar{r}) \left( \frac{\partial \bar{r}}{[\partial P_{jk}^a \ \partial P_{jk}^b]} \right)^{-1} \text{ with } \left( \frac{\partial \bar{r}}{[\partial P_{jk}^L \ \partial P_{jk}^L]} \right)^{-1} = \begin{bmatrix} N & 0 \\ 0 & N \end{bmatrix} \text{ and}$$

$$Var(\bar{r}) = \frac{1}{N^2} \begin{bmatrix} \sum_{n=1}^{N} \hat{\sigma}_n^2 & \sum_{n=1}^{N} \hat{\sigma}_n \hat{\tau}_n \\ \sum_{n=1}^{N} \hat{\sigma}_n \hat{\tau}_n & \sum_{n=1}^{N} \hat{\tau}_n^2 \end{bmatrix} \text{ where } \hat{\sigma}_n = \varpi_n \left\{ \left( \frac{p_{nj}}{p_{nk}} \right)^{1/2} - P_{jk}^a \right\} \text{ and } \hat{\tau}_n = \varpi_n \left\{ \left( \frac{p_{nk}}{p_{nj}} \right)^{1/2} - P_{jk}^b \right\}$$

which leads to

$$Var\begin{pmatrix} P_{jk}^a \\ P_{jk}^b \end{pmatrix} = \begin{bmatrix} \sum_{n=1}^{N} \hat{\sigma}_n^2 & \sum_{n=1}^{N} \hat{\sigma}_n \hat{\tau}_n \\ \sum_{n=1}^{N} \hat{\sigma}_n \hat{\tau}_n & \sum_{n=1}^{N} \hat{\tau}_n^2 \end{bmatrix} \quad (28)$$

Again, it is more useful to give the formula for variance of logarithm of an index.

Noting that $\ln P_{jk}^w = \ln P_{jk}^a - \ln P_{jk}^b$ along with application of variance of the sum and the delta method leads to

$$Var\left(\ln P_{jk}^W\right) = \sum_{n=1}^{N} \varpi_n^2 \left\{ \frac{1}{P_{jk}^a}\left(\frac{p_{nj}}{p_{nk}}\right)^{1/2} - \frac{1}{P_{jk}^b}\left(\frac{p_{nk}}{p_{nj}}\right)^{1/2} \right\}^2 \qquad (29)$$

### 5- Understanding and Interpreting Reliability Measures

One of the main products of the stochastic approach is a reliability measure (variance or standard error) for the estimated index. But how should such measures be interpreted? Three possible interpretations are discussed below.

*Variability over subsamples*: This interpretation is based on the theory and observation that the variance of an estimator can also be obtained from subsampling[10] therefore one can think of a computed variance or standard error as a measure of variability of an estimated index over subsamples. For example, for an index with standard error 0.1, the confidence interval $[\hat{P}_{jk} - 1.96 \times 0.1, \hat{P}_{jk} + 1.96 \times 0.1]$ gives an indication of how much the index changes over the subsamples[11]. In other words, it shows, how much on average the index varies if some items are randomly removed from the calculation of an index.

*Price dissimilarity measure*: Price dissimilarity metrics are measures of reliability of bilateral comparisons. They are used as a building block in elaborate spatial and temporal chaining methods (see e.g., Hill 1999, and Hajargasht et al. 2018, Diewert and Fox, 2020) and are also useful in deciding how to aggregate many price and quantity series into a smaller number of aggregates. Let $D_{jk}$ denote a dissimilarity measure or a distance metric for measuring the reliability of a bilateral comparison between a pair of $j$ and $k$. The smaller the value of $D_{jk}$ the more reliable the bilateral comparison is expected to be. Diewert (2009) has listed a set of axioms that a dissimilarity measure $D_{jk}$ needs to satisfy:

1) $D_{jj} = 0$
2) $D_{jk} \geq 0$
3) $D_{jk} = D_{kj}$
4) $D_{jk} = 0$ if $p_{nj} = \lambda p_{nk}$ for $n = 1,...., N$ and $\lambda > 0$
5) $D_{jk} = 0$ if and only if $p_{nj} = \lambda p_{nk}$ for $n = 1,...., N$ and $\lambda > 0$
6) $D_{jk}$ is invariant to the ordering of commodities.
7) $D_{jk}$ is invariant to the change in units of measurement.

---

[10] Subsampling is one of the resampling methods similar to the bootstrap used to approximate sampling distribution of an estimator. It differs from the bootstrap in (*i*) the resample size is smaller than the sample size (*ii*) resampling is done without replacement.
[11] 1.96 is the relevant critical value from the standard normal distribution table.

Diewert (2009) also introduced the following formulas as his preferred measures.

$$D^1_{jk} = \sum_{n=1}^{N} \left\{ \left( \frac{s_{nk} + s_{nj}}{2} \right) \left[ \left( \frac{p_{nj}}{p_{nk} P^F_{jk}} - 1 \right)^2 + \left( \frac{p_{nk} P^F_{jk}}{p_{nj}} - 1 \right)^2 \right] \right\}$$

$$D^2_{jk} = \sum_{n=1}^{N} \left\{ \left( \frac{s_{nk} + s_{nj}}{2} \right) \left[ \frac{p_{nj}}{p_{nk} P^F_{jk}} + \frac{p_{nk} P^F_{jk}}{p_{nj}} - 2 \right] \right\} \quad (30)$$

$$D^3_{jk} = \sum_{n=1}^{N} \left\{ \left( \frac{s_{nk} + s_{nj}}{2} \right) \left[ \ln \frac{p_{nj}}{p_{nk} P^T_{jk}} \right]^2 \right\} \text{ where } P^T_{jk} \text{ is a the Törnqvist price index.}$$

Here we argue that variance of logarithm of an ideal price index that satisfies time reversal test can be a good (perhaps a superior) dissimilarity measure. In particular, we propose the following measures:

$$D_{jk}^4 = Var\left(\ln P_{jk}^F\right) = \frac{1}{4}\sum_{n=1}^{N}\left\{s_{nk}\left(\frac{p_{nj}}{p_{nk}P_{jk}^L}-1\right) - s_{nj}\left(\frac{p_{nk}P_{jk}^P}{p_{nj}}-1\right)\right\}^2$$

$$D_{jk}^5 = Var\left(\ln P_{jk}^W\right) = \sum_{n=1}^{N}\varpi_n^2\left\{\frac{1}{P_{jk}^a}\left(\frac{p_{nj}}{p_{nk}}\right)^{1/2} - \frac{1}{P_{jk}^b}\left(\frac{p_{nk}}{p_{nj}}\right)^{1/2}\right\}^2 \qquad (31)$$

$$D_{jk}^6 = Var\left(\ln P_{jk}\right) = \sum_{n=1}^{N}\left\{\omega_n^2\left[\ln\frac{p_{nj}}{p_{nk}P_{jk}}\right]^2\right\} \text{ where } P_{jk} \text{ is one of the logarithmic indexes.}$$

It is straightforward to show that these measures satisfy all of Diewert's axioms. Our formulae have some resemblance to Diewert's measures but there are differences e.g., in (31) the weights are squared. The measures in (31) is likely to be superior to those in (30). First, the measures in (31) take into account the correlation between $p_{nj}/p_{nk}$ s and $p_{nk}/p_{nj}$ s. If the correlation is ignored, our formula resembles Diewert's measures more[12]. Second, the relevant properties of the underlying ideal indexes carry over to our proposed measures. For example, $D_{jk}^4$ and $D_{jk}^5$ in (31) satisfy the quantity reversal test but measures in (30) do not satisfy that[13]. Third, the new measures are always positive whereas the measures in (31) could become negative under extreme situations (e.g. in the presence of large negative weights). Fourth, introduction of weights in Diewert measures as $(s_{nk}+s_{nj})/2$ is one possibility whereas in our measure the weights emerge naturally. In fact, $(s_{nk}+s_{nj})/2$ could result in unreasonable large values for the measure in some situations. See Appendix-B for more explanations and a comparison between $D_{jk}^1$ and $D_{jk}^4$ based on a real data set.

---

[12] There will still be differences e.g., without correlation $D_{jk}^4 = \frac{1}{4}\sum_{n=1}^{N}\left\{s_{nk}^2\left(\frac{p_{nj}}{p_{nk}P_{jk}^L}-1\right)^2 + s_{nj}^2\left(\frac{p_{nk}P_{jk}^P}{p_{nj}}-1\right)^2\right\}$.

[13] The first measure in (31) has also the nice interpretation that it is equal to $\frac{1}{4}Var(\ln P_{jk}^L + \ln P_{jk}^P)$.

*Variability due to sampling*: Metrics such as those in (31) are also measures of the variability of an estimated index due to random sampling of $N$ items out of the population of all items[14]. Accordingly, one can use them to construct confidence intervals with standard interpretations. Such interpretations rely on the assumption of a simple random sampling where each item used in the computation of an index has the same chance to be in or out of the drawn sample. However, in practice often a stratified sampling design is used where items are classified into broad groups and random sampling is done within each group. In such cases, the variance relevant to this interpretation does not necessarily coincide with those in (31), and its calculation may require further information (e.g., within strata variances).

## 6- GEKS and its Variance

Now suppose that there are $M > 2$ countries (or time periods). In this case, the interest is on multilateral comparisons – comparison between all pairs of countries or periods – and the price index is expected to satisfy the transitivity property which requires $P_{jk} = P_{jl} \times P_{lk}$ for every $j, k, l \in 1,...,M$. There are several index number methods for constructing transitive multilateral price indexes (see e.g. Balk 1996, Hill 1997, and Diewert 1999). In this section we develop an appropriate stochastic approach to the GEKS[15]. The focus on GEKS is due to its critical role in the compilation of international

---

[14] Note that this differs from the first interpretation in that here we are considering variability due to the fact that we have sample rather than the entire population while there we were dealing with variability due to sampling out of the sample at hand. The value of these two are expected to coincide when the sample at hand represents the population well.

[15] The stochastic approach to these methods excluding the GEKS index has been discussed in Rao and Hajargasht (2016) among others. Cuthbert (2003) and Deaton (2012) have tried to compute standard

price and real expenditure comparisons from the ICP at the World Bank. Since the 2005 round, the ICP has been using the GEKS method at the recommendation of the Technical Advisory Group for the ICP. The GEKS has also been the aggregation method used for comparisons across countries in the EU and also among member countries of the OECD.

The GEKS method starts from the premise that the best comparison between a pair of countries/periods is a bilateral Fisher index (see e.g. Diewert 2013) and aims to construct transitive comparisons which are closest (according to a logarithmic least squares criterion) to those obtained from the bilateral Fisher index. The GEKS index is the solution to the following least squares problem

$$Min_{\ln P_{jk}} \sum_{j=1}^{M} \sum_{j=1}^{M} (\ln P_{jk} - \ln P_{jk}^{F})^2 \text{ subject to } P_{jk} = P_{jl}.P_{lk} \text{ for } j,k,l \in 1,...,M \tag{32}$$

It can be shown that this is equivalent to

$$Min_{\ln P_1^G,...,\ln P_M^G} \sum_{j=1}^{M} \sum_{j=1}^{M} (\ln P_j^G - \ln P_k^G - \ln P_{jk}^F)^2 \tag{33}$$

For this to have a unique solution, a normalization such as $P_M = 1$ is needed. In this case $P_{jM}^G$ and $P_{kM}^G$ are GEKS multilateral price indexes for country *k* and country *j* with country *M* as the base. The solution to (34) is equal to the estimates from regression

$$\ln P_{jk}^F = \ln P_{kM}^G - \ln P_{jM}^G + \varepsilon_{jk} \qquad j,k \in 1,...,M \tag{34}$$

It is straightforward to show that the GEKS index can be obtained as

$$\ln P_{jM}^G = \frac{1}{M} \sum_{k=1}^{M} [\ln P_{jk}^F + \ln P_{kM}^F] \tag{35}$$

---

errors for GEKS, but their estimates are not fully satisfactory. A stochastic approach to the Tornqvist based GEKS index is available in Rao and Selvanathan (1992).

To obtain variance of GEKS, one can use regression (34) along with an appropriate covariance matrix for $\varepsilon_{jk}$s. The obvious choice for this covariance matrix denoted by $\Omega$ is the matrix containing variances and covariances of all pairs of Fisher indexes. A variance sandwich formula $Var(\ln \mathbf{P}_{\_M}^G) = (\mathbf{J'J})^{-1}\mathbf{J'}\hat{\Omega}\mathbf{J}(\mathbf{J'J})^{-1}$ can then be used to obtain the variance of the GEKS index where $\mathbf{J}$ is a matrix of appropriately specified dummy variables and $\mathbf{P}_{\_M}^G = (P_{1M}^G, ......, P_{MM}^G)'$. An equivalent but easier approach is through direct application of the formula for variance of the sums to (35) or in matrix form

$$\ln P_{jM}^G = \frac{1}{M} \mathbf{i}'_{2M} \begin{bmatrix} \ln \mathbf{P}_{j\_}^F \\ \ln \mathbf{P}_{\_M}^F \end{bmatrix}$$

and $\quad Var(\ln P_{jM}^G) = \frac{1}{M^2} Var\left\{ \mathbf{i}'_{2M} \begin{bmatrix} \ln \mathbf{P}_{j\_}^F \\ \ln \mathbf{P}_{\_M}^F \end{bmatrix} \right\} = \frac{1}{M^2} \mathbf{i}'_{2M} Var\left\{ \begin{bmatrix} \ln \mathbf{P}_{j\_}^F \\ \ln \mathbf{P}_{\_M}^F \end{bmatrix} \right\} \mathbf{i}_{2M}$ (36)

where $\mathbf{P}_{j\_}^F = (P_{j1}^F, ......, P_{jM}^F)'$ and $\mathbf{P}_{\_M}^F = (P_{1M}^F, ......, P_{MM}^F)'$. Formula for variances of logarithm of the Fisher indexes was derived in Section 2, covariances can be obtained in a similar manner as

$$Cov\left(\ln P_{jk}^F, \ln P_{lm}^F\right) = \frac{1}{4} \sum_{n=1}^{N} \left\{ s_{nk}\left(\frac{p_{nj}}{p_{nk}P_{jk}^L} - 1\right) - s_{nj}\left(\frac{p_{nk}P_{jk}^P}{p_{nj}} - 1\right) \right\} \left\{ s_{nl}\left(\frac{p_{nl}}{p_{nm}P_{lm}^L} - 1\right) - s_{nm}\left(\frac{p_{nm}P_{lm}^P}{p_{nl}} - 1\right) \right\} \quad (37)$$

Using (16), (37) along with (36), variance of GEKS can be easily estimated.

### 7- Application to 2017 ICP Data

In this section, 2017 ICP data is used to estimate all the indexes mentioned above along with their reliability measures. ICP collects price data on a large number of items from the participating countries (173 countries in 2017 round). In the first stage, collected prices of the items are aggregated into basic headings (154 of them in 2017) without weights[16]. In the second stage, basic heading (BH) prices are aggregated upwards with expenditure weights to the GDP level or other major aggregates. We estimate GDP level PPPs for all countries using price and expenditure data provided by ICP on 154 BHs in 173 countries. See Rao (2013) for further details on the framework and the steps involved in price and real expenditure comparisons.

As the ICP publishes price comparisons for each country with USA as the base country, we first compute the bilateral Fisher price index for each country with US as numeraire. From a stochastic point of view, the standard errors of the computed Log-Fisher indexes are of interest[17]. They range between 0.031 to 0.187 where the smallest standard errors belong to countries that are "similar" to US (e.g., Canada, Australia, Scandinavian countries, ….) and the largest standard errors belong to countries considered "dissimilar" to US such as Sierra Leone, Tajikistan and Ukraine. All the estimated standard errors can be seen[18] from Figure-1.

Figure-1 compares nonparametric bootstrap standard errors[19] with standard errors calculated based on Formula (16). Note that these bootstrap standard errors are not based on any particular stochastic specification and the sampling is done over the weights as well. The almost identical numbers suggests that our model-based standard errors (16) are robust to whatever the true underlying stochastic model is.

---

[16] The details of the PPP compilation process can be found in World Bank (2020).
[17] Note that
[18] For the name of countries see Country Codes (worldbank.org).
[19] The bootstrap estimates are based on 2000 resampling.

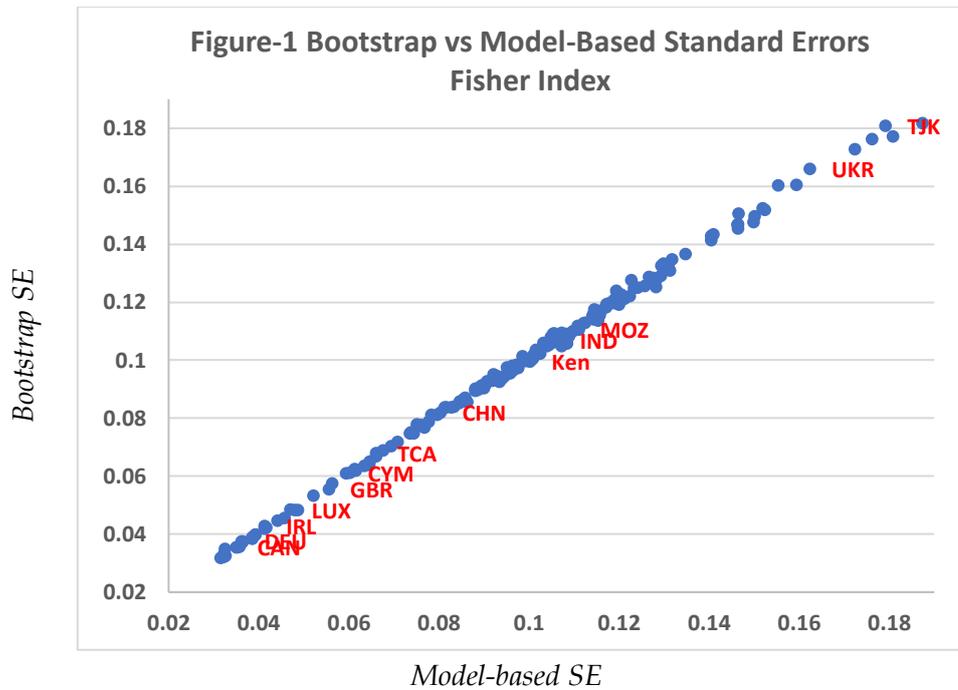

To compare other notable indexes with the Fisher index, Tabel-1 presents the average absolute value of the difference of Log of each index from Log of Fisher index.

| Tabel-1: Percentage Difference of various Indexes from Fisher Index | | | | | |
|---|---|---|---|---|---|
| Törnq | Lasp | Paasch | CPD | SV | Walsh |
| 1.85 | 14.64 | 14.64 | 3.02 | 2.12 | 2.23 |

Törnqvist is the closest to Fisher with a 1.85% difference on average. In second place is the Sato-Vartia index with 2.12% and Walsh is next with a 2.23% difference. CPD has the highest deviation excluding the Laspeyres and Paasche.

Figure-2 depicts the percentage difference between these indexes with the Fisher index. As it can be seen, the differences do not appear systematic in the sense that the points are scattered relatively evenly around zero at every level of Log-Fisher index. As expected, Törnqvist has the lowest overall deviation from the Fisher index. We see a few large negative Log-differences for the CPD index.

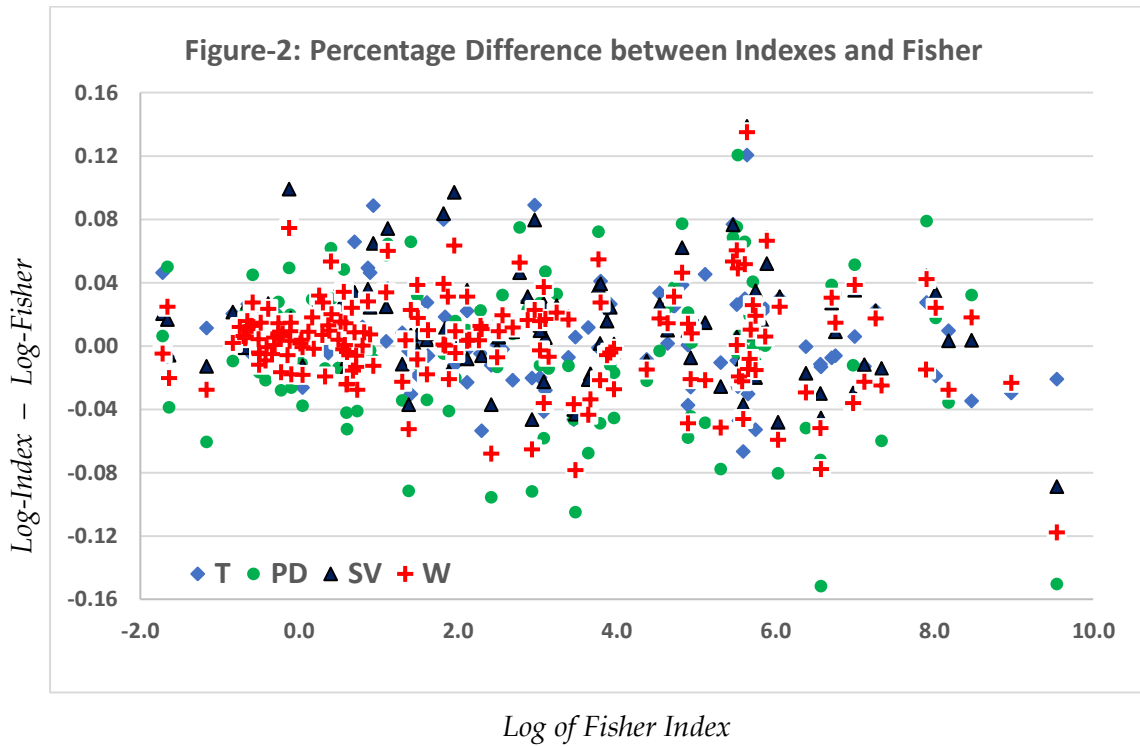

Figure-2 depicts standard errors of other indexes versus the standard error of the Fisher index. In general, standard errors from Log-Fisher appear to be smaller than those from other indexes. Again, Törnqvist is the closest while Sato-Vartia and Walsh have more or less the same level of deviation from the Fisher index. Again, standard errors from bilateral CPD are generally larger. It also appears that with the increase in the standard errors of Log-Fisher index, the difference between standard error of the indexes with those from the Log-Fisher index increases.

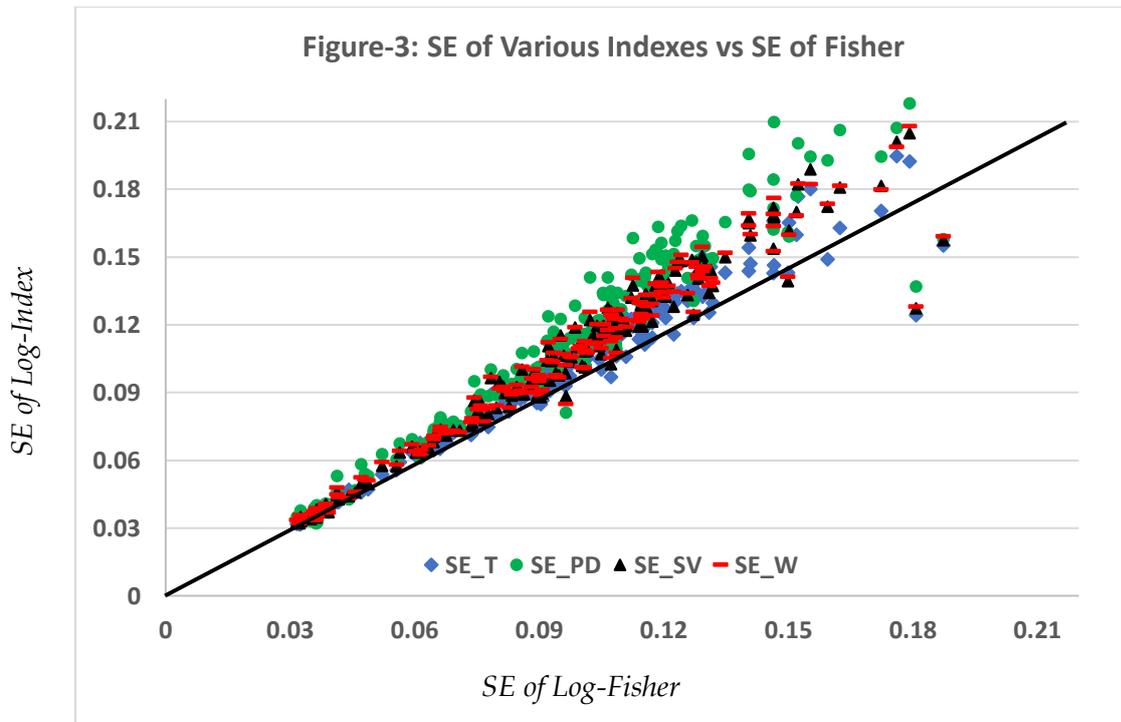

Now we use the method discussed in Section 6 to compute the multilateral GEKS index and its reliability measure for all countries with US as numeraire. One question of interest is how different the multilateral GEKS and the bilateral Fisher indexes are. The difference is depicted in Figure-4 According to our estimates, the difference between GEKS and the Fisher index for most countries is less than 5% in absolute value. There are some countries (e.g., IR, VGB, LUX, CYM, TCA) where GEKS is substantially smaller than the bilateral Fisher index. These are small countries with a large negative expenditure on net exports. Interestingly, China is also among countries with a high level of difference (around 8%). Figure-4 shows the percentage difference between GEKS and Fisher versus the standard error of LogFisher. Another point that this graph reveals is that there are countries that are quite similar to US such as Canada and Japan (reflected in their small bilateral standard errors) where the GEKS index is noticeably bigger than the bilateral Fisher. This could potentially mean that going from bilateral to a multilateral comparison could make the comparison less reliable for these group of countries.

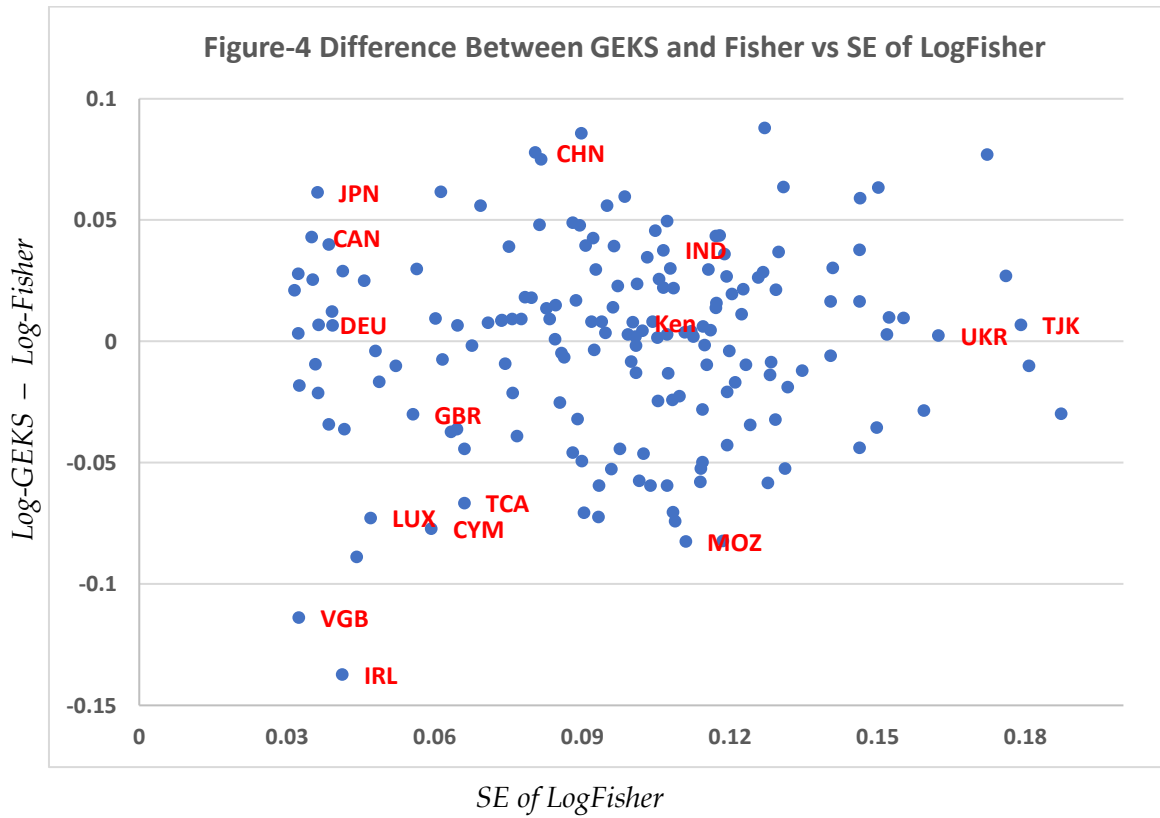

To investigate this in more detail we have created Figure-5 which compares standard error of the GEKS and Fisher indexes for each country. Here are a few points that can be made based on this figure: (*i*) overall the GEKS standard errors tend to be smaller. This could mean that going from a bilateral to a multilateral comparison increases the precision of the estimated indexes perhaps due to inclusion of countries in the mix that are more similar both to themselves and to US. (*ii*) There are exceptions though, some of the countries that are considered similar to US (e.g., Scandinavian countries, Australia) end up with higher standard errors. This could be interpreted as the negative impact of brining less similar countries into the mix for these countries.

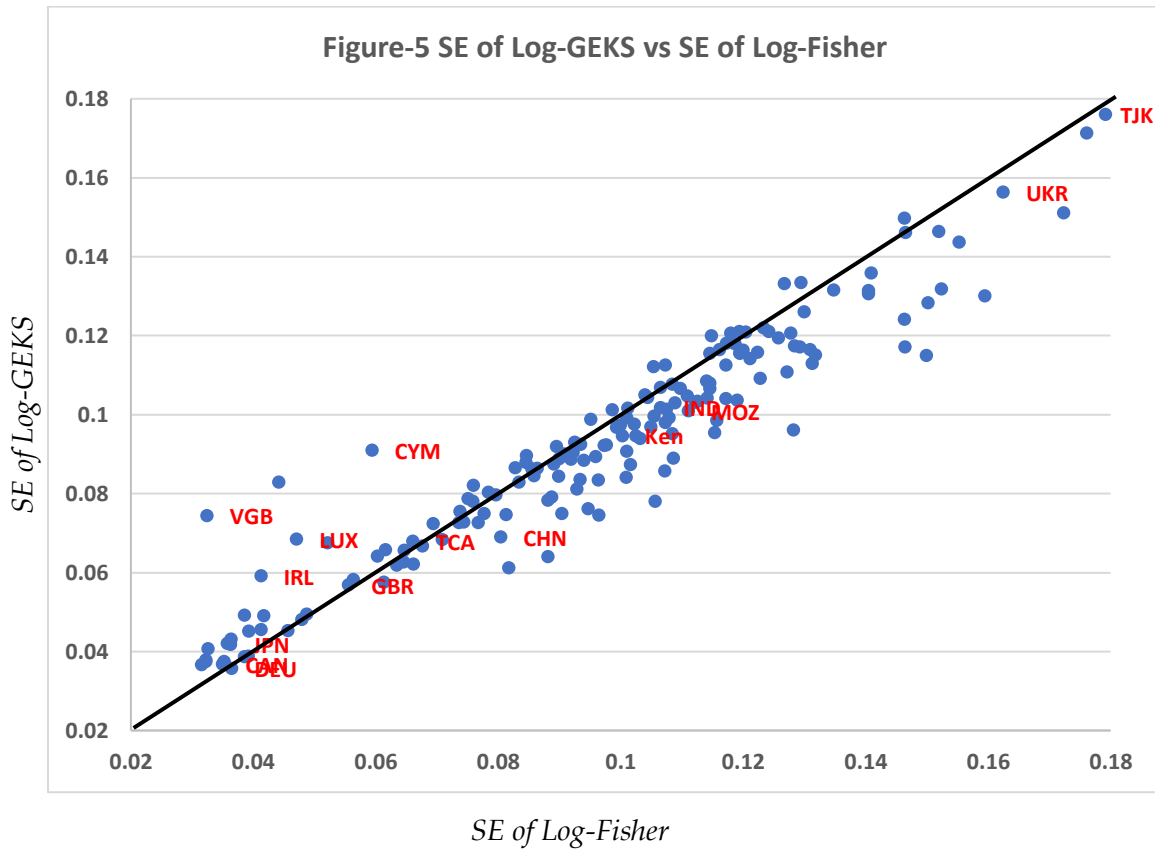

*SE of Log-Fisher*

(*iii*) Standard error of log-GEKS is substantially higher for the countries with large negative net exports consistent with large difference between Log-GEKS and Log-Fisher for them. It appears that for these countries going form a bilateral to a multilateral comparison decreases the precision indicating that for these countries the GEKS estimates should be trusted less. (*iv*) For China, the multilateral SE is substantially smaller suggesting that GEKS is a more reliable index than the bilateral Fisher. This could potentially mean that for China, the GEKS estimate should be trusted more.

## 8- Conclusion

Price indexes (CPIs and PPPs) are often published without any indicator of the uncertainty around them. This paper demonstrates that sensible and principled reliability measures for ideal bilateral and multilateral price indexes can be estimated.

It also clarifies how they should be interpreted. The empirical results shows that the estimated measures are robust and can say a lot about the reliability of bilateral or multilateral price comparisons. For example, they can show that going from a bilateral to a multilateral index makes the price comparison less reliable for countries that are more similar in terms of price and quantities to the base county (USA) and more reliable for countries such as China that are less similar to the base country. The proposed reliability measures can also be interpreted as dissimilarity measures and used as a building block in elaborate spatial and temporal chaining methods for making better multilateral comparisons. Given the robustness and usefulness of such measures, national and international agencies are highly recommended to publish reliability measures along with their estimated indexes.

National accounts in constant prices and in PPPs are widely used for measuring welfare across time and in cross country comparisons but it is not clear how certain the published figures are. The stochastic approach of this paper can be extended to compute principled reliability measures for national accounts as well. The approach can also be extended to prominent multiperiod index numbers such as the fixed base, the chained base and the rolling GEKS.

**Appendix-A: "Equivalence" of Two Stochastic Approaches**

As mentioned in Section 2, there are at least two approaches to specifying a stochastic model that can generate a price index. In the approach described in Section 2, the stochastic model is specified in terms of price ratios. In the other approach, it is assumed that each price $p_{nj}$ can approximately be written as product of two terms i.e. $p_{nj} = P_j \alpha_n$ where $P_j$ is the general price level at location/period $j$ and $\alpha_n$ is the part specific to a particular item. This formula is sometimes referred to as the *law of one price*. Often an error term is appended, and a stochastic model is specified as

$$p_{nj} = P_j \alpha_n e_{nj} \text{ with } E(e_{nj}) = 1 \qquad \text{(A-1)}$$

or a slight modification of it

$$\frac{p_{nj}}{P_j \alpha_n} - 1 = \varepsilon_{nj} \text{ with } E(\varepsilon_{nj}) = 0 \qquad \text{(A-2)}$$

The purpose of this section is to show that with the right choice of weights and appropriate assumptions about the error term, the two stochastic approaches produce the same indexes and remarkably the same variance despite differing specifications and weights. We first consider the Laspeyres, Paasche and Walsh indexes and later discuss logarithmic indexes.

Let there be two countries/periods of $j$ and $k$ with the $k$-th country/period as base (i.e. $P_k = 1$), then our stochastic equations can be written as

$$\begin{cases} \dfrac{p_{nj}}{P_{jk}\alpha_n} - 1 = \varepsilon_{nj} \\ \dfrac{p_{nk}}{\alpha_n} - 1 = \varepsilon_{nk} \end{cases} \quad n = 1,...., N \tag{A-3}$$

Here we show that (A-3) can be estimated using the nonlinear-nonadditive regression framework presented in Rao and Hajargasht (2016). First define $2N \times 1$ vector $\mathbf{r}$ as follows:

$$\mathbf{r} = \begin{bmatrix} \dfrac{p_{nj}}{P_{jk}\alpha_n} - 1 \\ \dfrac{p_{nk}}{\alpha_n} - 1 \end{bmatrix}_{n=1,...,N} \tag{A-4}$$

It has been shown that one can estimate (A-4) using a weighted method of moments[20] $\mathbf{R}'\mathbf{W}\mathbf{r} = \mathbf{0}$ where $\mathbf{W}_{2N \times 2N}$ is a diagonal matrix with appropriate weights in its diagonal. The optimal (minimum variance) choice for $\mathbf{R}$ is

$$\mathbf{R}^{*}_{(N+1) \times 2N}{}' = E\left(\dfrac{\partial \mathbf{r}}{\partial [P_{jk}, \boldsymbol{\alpha}]'}\right) = Diag\begin{pmatrix} -1/P_{jk} & 0 \\ 0 & -1/\boldsymbol{\alpha} \end{pmatrix}\begin{pmatrix} \mathbf{i}'_N & 0 \\ \mathbf{I}_N & \mathbf{I}_N \end{pmatrix} \tag{A-5}$$

where $Diag(.)$ means a diagonal matrix, $\mathbf{i}'_N = [1,...,1]$, $\mathbf{I}_N$ is the identity matrix of size $N$ and $1/\boldsymbol{\alpha}' = [1/\alpha_1,....,1/\alpha_N]$. This choice of $\mathbf{R}$ does not lead to the Laspeyres or Paasche indexes but a slight modification of it i.e.

$$\mathbf{R}'_{(N+1) \times 2N} = Diag\begin{pmatrix} -1/P_{jk} & 0 \\ 0 & -1/\boldsymbol{\alpha} \end{pmatrix}\begin{pmatrix} \boldsymbol{\alpha}'_N & 0 \\ \mathbf{I}_N & \mathbf{I}_N \end{pmatrix} \text{ with } \mathbf{W}_{2N \times 2N} = Diag\begin{pmatrix} \mathbf{q}_{nk} & 0 \\ 0 & \mathbf{q}_{nk} \end{pmatrix} \tag{A-6}$$

with $\mathbf{q}_{nk} = [q_{1k},....,q_{Nk}]$ leads to the Laspeyres index as it is shown below.

---

[20] This is a variation of the method of moments of Section 2.

$$\mathbf{R'Wr} = \mathbf{0} = \begin{cases} \sum_{n=1}^{N} \dfrac{-\alpha_n}{P_{jk}} q_{nk} \left( \dfrac{p_{nj}}{P_{jk}\alpha_n} - 1 \right) = 0 \\ q_{nk} \left( \dfrac{-1}{\alpha_n} \right) \left( \dfrac{p_{nj}}{P_{jk}\alpha_n} - 1 \right) + q_{nk} \left( \dfrac{-1}{\alpha_n} \right) \left( \dfrac{p_{nk}}{\alpha_n} - 1 \right) = 0 \quad n = 1,...,N \end{cases} \quad (A\text{-}7)$$

using some algebraic manipulations, we can show that

$$\left. \begin{aligned} \dfrac{1}{P_{jk}} &= \dfrac{\sum_{n=1}^{N} q_{nk}\alpha_n}{\sum_{n=1}^{N} q_{nk} p_{nj}} \\ \alpha_n &= \dfrac{p_{nj}}{2P_{jk}} + \dfrac{p_{nk}}{2} \end{aligned} \right\} \Rightarrow P_{jk} = \dfrac{\sum_{n=1}^{N} p_{nj} q_{nk}}{\sum_{n=1}^{N} p_{nk} q_{nk}} \quad (A\text{-}8)$$

The Paasche and Walsh indexes can be obtained in a similar fashion by using

$$\mathbf{W}_P = Diag\begin{pmatrix} \mathbf{q}_{nj} & 0 \\ 0 & \mathbf{q}_{nj} \end{pmatrix} \text{ and } \mathbf{W}_W = Diag\begin{pmatrix} \sqrt{\mathbf{q}_{nj}\mathbf{q}_{nk}} & 0 \\ 0 & \sqrt{\mathbf{q}_{nj}\mathbf{q}_{nk}} \end{pmatrix} \text{ respectively.}$$

One of the virtues of the stochastic approach is the possibility of obtaining variances or reliability measures. Variance of the estimated indexes can be obtained using the sandwich formula

$$Var(\hat{\alpha}_1,...,\hat{\alpha}_N, \hat{P}_{jk}) = \left[\mathbf{R'WR}^*\right]^{-1} \mathbf{R'W\Omega WR} \left[\mathbf{R'WR}^*\right]^{-1} \quad (A\text{-}9)$$

This formula and some tedious algebra can be used to prove that the variance of $\hat{P}_{jk}$ obtained from this equation is equal to the variance obtained in Section 3. Here we use an indirect less tedious approach to prove this. First, using definition of the Laspeyres index and the delta method

$$Var\left(\ln P_{jk}^L\right) = \dfrac{Var\left(\sum_{n=1}^{N} q_{nk} p_{nj}\right)}{\left(\sum_{n=1}^{N} q_{nk} p_{nj}\right)^2} + \dfrac{Var\left(\sum_{n=1}^{N} q_{nk} p_{nk}\right)}{\left(\sum_{n=1}^{N} q_{nk} p_{nk}\right)^2} - 2\dfrac{Cov\left(\sum_{n=1}^{N} q_{nk} p_{nj}, \sum_{n=1}^{N} q_{nk} p_{nk}\right)}{\left(\sum_{n=1}^{N} q_{nk} p_{nj}\right)\left(\sum_{n=1}^{N} q_{nk} p_{nk}\right)} \quad (A\text{-}10)$$

Since covariances across non-identical items are assumed zero, using equation (A-3) and suppressing "^"s, we have $Var(p_{nj}) = \alpha_n^2 P_{jk}^2 \sigma_{nj}^2$, $Var(p_{nk}) = \alpha_n^2 \sigma_{nk}^2$, $Var(p_{nj}) = \alpha_n^2 P_{jk} \sigma_{n,jk}$ and

$$Var\left(\ln P_{jk}^L\right) = \sum_{n=1}^{N} q_{nk}^2 \left\{ \frac{\left(P_{jk}\alpha_n\sigma_{nj}\right)^2}{\left(\sum_{n=1}^{N} p_{nj}q_{nk}\right)^2} + \frac{\left(\alpha_n\sigma_{nk}\right)^2}{\left(\sum_{n=1}^{N} p_{nk}q_{nk}\right)^2} - \frac{2P_{jk}\alpha_n^2\sigma_{n,k}}{\left(\sum_{n=1}^{N} p_{nj}q_{nk}\right)\left(\sum_{n=1}^{N} p_{nk}q_{nk}\right)} \right\} \quad \text{(A-11)}$$

Using the estimates $\hat{\sigma}_{nj}^2 = \hat{\varepsilon}_{nj}^2, \hat{\sigma}_{nj}^2 = \hat{\varepsilon}_{nk}^2$ and $\hat{\sigma}_{n,jk} = \hat{\varepsilon}_{nj}\hat{\varepsilon}_{nk}$ we can write

$$Var\left(\ln P_{jk}^L\right) = \sum_{n=1}^{N} q_{nk} \left\{ \frac{\left(P_{jk}^L \alpha_N \left(\frac{p_{nj}}{P_{jk}^L \alpha_N} - 1\right)\right)}{\left(\sum_{n=1}^{N} p_{nj}q_{nk}\right)} - \frac{\left(\alpha_N \left(\frac{p_{nk}}{\alpha_N} - 1\right)\right)}{\left(\sum_{n=1}^{N} p_{nk}q_{nk}\right)} \right\}^2$$

$$= \sum_{n=1}^{N} \left\{ \frac{p_{nk}q_{nk}}{\left(\sum_{n=1}^{N} p_{nk}q_{nk}\right)} \left[ \alpha_N \left(\frac{p_{nj}}{p_{nk}P_{jk}^L \alpha_N} - \frac{1}{p_{nk}}\right) - \alpha_N \left(\frac{1}{\alpha_N} - \frac{1}{p_{nk}}\right) \right] \right\}^2 \quad \text{(A-12)}$$

$$= \sum_{n=1}^{N} \left\{ s_{nk} \left(\frac{p_{nj}}{p_{nk}P_{jk}^L} - 1\right) \right\}^2$$

This is exactly the formula we derived in Section 3 for variance of the logarithm of the Laspeyres index. The same can be proven for the Paasche and Walsh indexes.

In what follows we demonstrate the PD approach and its equivalence for logarithmic indexes. Consider the logarithmic form of equation (A-1)

$$\begin{aligned}\ln p_{nj} &= \ln P_j + \delta_n + \varepsilon_{nj} \\ \ln p_{nj} &= \ln P_k + \delta_n + \varepsilon_{nk}\end{aligned} \quad \text{(A-13)}$$

with the following assumptions on the error term

$$\begin{aligned} E(\varepsilon_{nj}) &= 0, \; E(\varepsilon_{nk}) = 0 \\ Var(\varepsilon_{nj}) &= \sigma_{nj}^2, Var(\varepsilon_{nk}) = \sigma_{nk}^2 \\ Cov(\varepsilon_{nj}, \varepsilon_{nk}) &= \sigma_{n,jk} \end{aligned} \quad \text{(A-14)}$$

Then the model can be estimated using weighted least squares (or method of

moments):

$$Min_{\{P_{jk},\delta_n\}} \sum_{n=1}^{N} \left\{ \psi_{nj} \left( \ln p_{nj} - \ln P_{jk} - \delta_n \right)^2 + \psi_{nk} \left( \ln p_{nk} - \delta_n \right)^2 \right\} \quad \text{(A-15)}$$

It is easy to show that with the following choices of weights, we obtain PD, Törnqvist and Sato-Vartia indexes.

PD: $\psi_{nj} = s_{nj}$ and $\psi_{nk} = s_{nk}$

Törnqvist: $\psi_{nj} = \psi_{nk} = \omega_n^A = (s_{nj} + s_{nk})/2$  (A-16)

Sato-Vartia: $\psi_{nj} = \psi_{nk} = \omega_n^L = \dfrac{L(s_{nj}, s_{nk})}{\sum_{n=1}^{N} L(s_{nj}, s_{nk})}$

Now we show that the variance obtained from both approaches are equal when the appropriate assumptions i.e. (A-14) is made about the error term.

$$Var\left(\ln P_{jk}\right) = Var\left(\sum_{n=1}^{N} \omega_n \left(\ln p_{nj} - \ln p_{nk}\right)\right) = \sum_{n=1}^{N} \omega_n^2 (\sigma_{nj}^2 + \sigma_{nk}^2 - 2\sigma_{n,jk}) \quad \text{(A-17)}$$

Assuming robust estimators for variances we have

$$\sum_{n=1}^{N} \omega_n^2 (\hat{\sigma}_{nj}^2 + \hat{\sigma}_{nk}^2 - 2\hat{\sigma}_{n,jk}) = \sum_{n=1}^{N} \omega_n^2 (\hat{\varepsilon}_{nj}^2 + \hat{\varepsilon}_{nk}^2 - 2\hat{\varepsilon}_{nj}\hat{\varepsilon}_{nk}) \quad \text{(A-18)}$$

This can be further simplified to

$$Var\left(\ln P_{jk}\right) = \sum_{n=1}^{N} \omega_n^2 (\hat{\varepsilon}_{nj} - \hat{\varepsilon}_{nk})^2 = \sum_{n=1}^{N} \omega_n^2 (\ln p_{nj} - \ln P_{jk} - \hat{\delta}_n - \ln p_{nk} + \hat{\delta}_n)^2 = \sum_{n=1}^{N} \omega_n^2 \left(\ln \dfrac{p_{nj}}{p_{nk}} - \ln P_{jk}\right)^2 \quad \text{(A-19)}$$

which is the same formula derived in Section 4.

**Appendix-B: A Comparison of Dissimilarity Measures**

Here we compute Diewert's $D_{jk}^1$ measure in (30) and our proposed dissimilarity measure $D_{jk}^4$ in (31) for 173 countries with respect to USA using 2017 round of ICP[21].

---

[21] These two measures have very different scales, to make the numbers more comparable across the two measures, we have multiplied our measure by 150.

The results have been depicted in Figure-A. The first thing to note is that the two measures are highly correlated (Correlation Coeff = 0.79) but there are also significant differences. Second, the two measures follow each other better at lower levels of dissimilarity (e.g. for the first half of dissimilarity measures, Correlation Coeff = 0.92) but at higher levels, $D_{jk}^1$ is generally larger than $D_{jk}^4$. Third, there are cases where the two measures are significantly different (e.g. NPL, SAU, KWT, CHN and KOR).

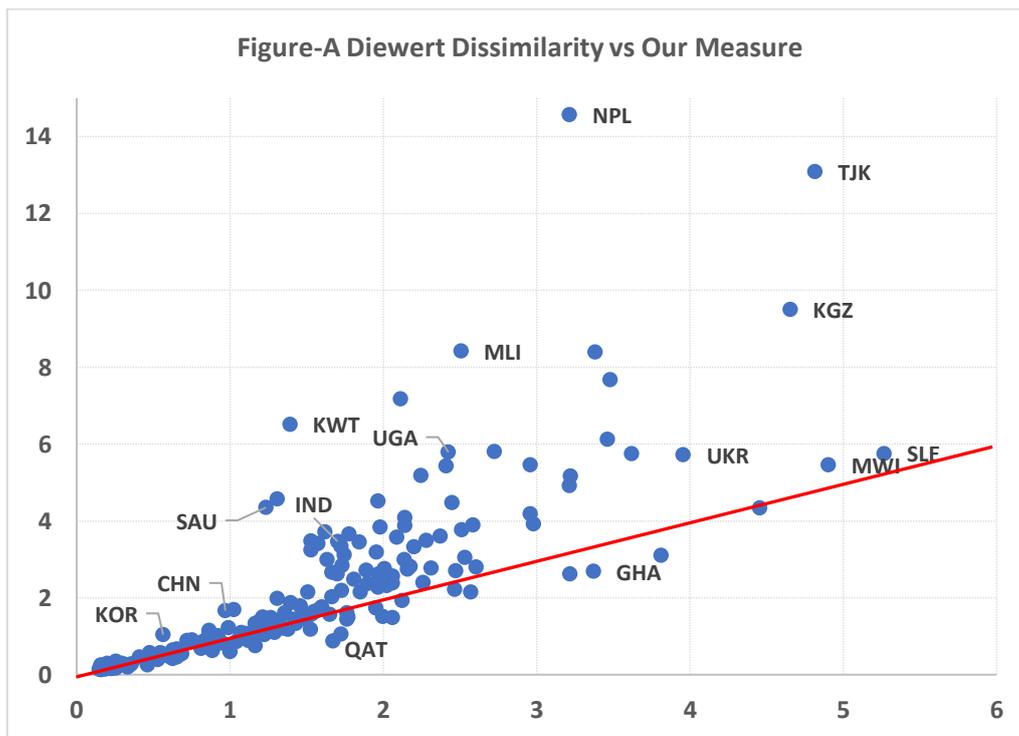

Figure-A Diewert Dissimilarity vs Our Measure

Further investigation shows that there are situations where $D_{jk}^1$ is driven by an item even if the weight of that item is very small. For example, in case of Nepal, $D_{jk}^1$ is equal to 14.57 but 7.89 of this is due to item $n=43$. For this item, the price ratio (0.63) is substantially different from the Fisher index (38.94) but the weight of the item is small i.e. 0.00089. In $D_{jk}^1$ this weight is averaged with the much bigger weight of the item from USA (0.0033785), and then is multiplied by $\left[\left(p_{nj}/p_{nk}P_{jk}^F-1\right)^2+\left(p_{nk}P_{jk}^F/p_{nj}-1\right)^2\right]$

which create the large value of 7.89. In fact, consider an extreme situation where $s_{nk}$ for item $n$ is large and $p_{nj}/p_{nk}$ is substantially different from the estimated index and $s_{nj} = 0$. We would not expect this item to have a substantial impact on dissimilarity measure since $s_{nj} = 0$ but one can create artificial examples where this item contributes to more than 99% of the estimated $D_{jk}^1$. Our measure $D_{jk}^4$ is not prone to this issue to the same extent, the contribution of $n = 43$ is only 0.079 in $D_{jk}^4 = 3.21$.